\documentclass[aps,twocolumn,showpacs]{revtex4}
\usepackage{graphicx}
\usepackage{amsfonts}
\usepackage{amssymb}
\usepackage{amsbsy}
\usepackage{amsmath}
\usepackage{mathrsfs}
\usepackage{latexsym}
\usepackage{natbib}
\usepackage{bm}
\usepackage{subfigure}
\usepackage{color}

\def\kr{\kappa}

\def\ksg{\mathrm{\varkappa}}

\def\rs{r_s}
\def\rstar{r_{\star}}
\def\scriplus{\mathscr{I}^{+}}
\def\scriminus{\mathscr{I}^{-}}

\def\observerminus{\mathbb{O}^{-}}
\def\observerplus{\mathbb{O}^{+}}

\def\lstar{l_{\star}}

\begin{document}

\title{The Hawking effect is short-lived in polymer quantization}

\author{Subhajit Barman}
\email{sb12ip007@iiserkol.ac.in}

\author{Golam Mortuza Hossain}
\email{ghossain@iiserkol.ac.in}

\author{Chiranjeeb Singha}
\email{cs12ip026@iiserkol.ac.in}

\affiliation{ Department of Physical Sciences, 
Indian Institute of Science Education and Research Kolkata,
Mohanpur - 741 246, WB, India }
 

\date{\today}

\begin{abstract}

It is widely believed that the Hawking effect might hold clues to the possible, 
yet unknown, trans-Planckian physics. On the other hand, one could ask whether 
the effect itself might be altered by such trans-Planckian physics. We seek an 
answer to this question within a framework where matter field is quantized using 
polymer quantization, a canonical quantization technique employed in loop 
quantum gravity. We provide an exact derivation of the Hawking effect using 
canonical formulation by introducing a set of near-null coordinates which allows 
one to overcome the challenges posed by a Hamiltonian-based derivation of the 
Hawking effect. Subsequently, we show that in polymer quantization the Hawking 
effect is short-lived and it eventually disappears for an asymptotic future 
observer. Such an observer finds the duration of the Hawking effect to be few 
milliseconds for a solar mass black hole whereas it is few years for an 
ultra-massive black hole. Consequently, it provides a new way to resolve the 
so-called information loss paradox.

\end{abstract}

\maketitle

\emph{Introduction.}--
The \emph{Hawking effect} \cite{hawking1975} continues to be an enigma in modern 
physics where an asymptotic future observer experiences a thermal emission, 
rather unexpectedly, emanating from a classical black hole. In statistical 
physics, the thermal emissions are known to arise from systems having large 
number of microscopic degrees of freedom. However, to describe black holes which 
are solutions of Einstein's general relativity, only a few parameters are 
required. This perplexing property suggests that the Hawking effect might hold 
the key in understanding possible microstates of a black hole. These states are 
expected to arise from a possible, yet unknown, quantum theory of gravity and 
have been pursued extensively in different contexts 
\cite{Hawking:review.papers}.

It is well known that all prominent derivations of the Hawking effect rely  on 
the properties of the trans-Planckian frequencies one way or other. On the other 
hand, it's widely expected that our current understanding of trans-Planckian 
physics would need to be modified in order to tame the plaguing ultraviolet 
divergences. Therefore, one is led to ask whether the Hawking effect itself 
could survive these expected trans-Planckian modifications 
\cite{cutoff:3article}. Besides, the evaporation of a black hole through Hawking 
radiation gives rise to the so-called information loss 
paradox \cite{Information:3article,Unruh:2017uaw} which, according to the 
popular school of thought, threatens \emph{unitarity}, a key pillar of quantum 
theory (see also \cite{Maudlin:2017lye}).

We seek an answer to the question within the framework of polymer quantization 
of matter field in the Schwarzschild geometry which is formed through a 
collapsing shell of matter. Polymer or loop quantization 
\cite{Polymer:loop2papers} is a canonical quantization technique which is 
employed in loop quantum gravity \cite{Loop:3papers}. This quantization comes 
with a new \emph{length scale} which would correspond to the Planck length in 
full quantum gravity. However, here we would employ this quantization only for 
the matter sector and treat the spacetime geometry as the classical entity 
\cite{curved:3books}, as done for the standard derivation of the Hawking effect. 
Similar studies in the context of the Unruh effect \cite{Unruh:3papers} has 
indicated significant modification \cite{Hossain:2014fma,Hossain:2papers}.

It turns out that there are major hurdles in pursuing a Hamiltonian-based 
derivation of the Hawking effect. The key reason behind these hurdles is the 
fact that thermal characteristic of the Hawking quanta is realized using the 
relation between the modes that leave past null infinity as ingoing null rays 
and the modes that arrive at future null infinity as outgoing null rays. 
Expectedly, the usage of the advanced and retarded \emph{null coordinates} 
rather than the regular Schwarzschild coordinates, forms a crucial backbone for 
the standard derivation of the Hawking effect. However, these null coordinates 
do not lead to a true Hamiltonian that describes evolution of the relevant modes 
(see also \cite{Null:hamiltonian}). In an earlier such attempt by Melnikov and 
Weinstein \cite{Melnikov:2001ex} who used Lema\^itre coordinates, the Hawking 
effect is understood indirectly through the property of the Green's function. To 
the best of our knowledge an exact derivation of the thermal spectrum for 
Hawking radiation using Hamiltonian formulation is still lacking.

\emph{Schwarzschild spacetime.}--
Let us consider a Schwarzschild black hole which is formed after the collapse 
of a matter shell. The corresponding metric is
\begin{equation}\label{SchwarzschildMetric0}
ds^2 = - \Omega dt^2 + \Omega^{-1} dr^2 
+ r^2 d\theta^2 + r^2 \sin\theta^2 d\phi^2 ~,
\end{equation}
where $\Omega = \left(1- r_s /r\right)$. Here we have chosen \emph{natural 
units} such that $c=\hbar=1$ and the Schwarzschild radius $r_s = 2 G M$.
If one defines the so-called \emph{tortoise coordinate} $\rstar$ such that 
$d\rstar = \Omega^{-1} dr$, then $t-r$ plane of the Schwarzschild geometry 
becomes \emph{conformally flat}. By a suitable choice of constant of 
integration, $\rstar$ can be written as $\rstar = r + r_s \ln 
\left(r/r_{s}-1\right)$. For later convenience, we define the \emph{advanced} 
and \emph{retarded} null coordinates $v = t + \rstar$ and $u = t - \rstar$ 
respectively.

In addition to the collapsing shell of matter, we consider a minimally coupled, 
massless scalar field $\Phi(x)$ to represent the Hawking quanta 
\cite{hawking1975}, and which is governed by the action $S_{\Phi} = \int d^{4}x 
\left[ -\frac{1}{2} \sqrt{-g} g^{\mu \nu} \nabla_{\mu}\Phi(x) 
\nabla_{\nu}\Phi(x) \right]$. For an observer at \emph{past null infinity} 
$\scriminus$, the scalar field operator can be expressed as 
\begin{equation}\label{ScalarFieldPastOperator}
\hat{\Phi}(x) = \sum_{\omega} \left[ 
{f}_{\omega} \hat{a}_{\omega} + {f}^{*}_{\omega} \hat{a}^{\dagger}_{\omega} 
\right] ~,
\end{equation}
where the set of \emph{ingoing} field solutions $\{{f}_{\omega}\}$ 
forms a complete family on $\scriminus$ along with the inner product
$(-i/2)\int_{S}d\Sigma^{a} \left({f}_{\omega} \nabla_a {f}^{*}_{\omega'} -
{f}^{*}_{\omega'} \nabla_a {f}_{\omega} \right) = \delta_{\omega\omega'}$ 
where $S=\scriminus$.
In order to render the inner product \emph{positive definite}, only positive 
frequency  modes, with respect to a canonical affine parameter along 
$\scriminus$, are chosen. These modes can be written as
\begin{equation}\label{IngoingSolution}
{f}_{\omega}(v) = \frac{1}{\sqrt{2\pi\omega}} ~ r^{-1}~ 
e^{-i\omega v} ~Y_{lm}(\theta,\phi) ~,
\end{equation}
where $Y_{lm}(\theta,\phi)$ are \emph{spherical harmonics}. The creation and 
annihilation operators are $\hat{a}^{\dagger}_{\omega}$ and $\hat{a}_{\omega}$ 
respectively. The \emph{vacuum} state $|0_{-}\rangle$ is defined as 
$\hat{a}_{\omega}~|0_{-}\rangle = 0$.
Similarly for a future observer we can express $\hat{\Phi}(x)$ as
\begin{equation}\label{ScalarFieldFutureOperator}
\hat{\Phi}(x) = 
\sum_{\omega} \left[ {p}_{\omega} \hat{b}_{\omega} + {p}^{*}_{\omega} 
\hat{b}^{\dagger}_{\omega} \right]
+
\sum_{\omega} \left[ {q}_{\omega} \hat{c}_{\omega} + {q}^{*}_{\omega} 
\hat{c}^{\dagger}_{\omega} \right]
~,
\end{equation}
where field solutions ${p}_{\omega}(u) = \tfrac{1}{\sqrt{2\pi\omega}} ~ r^{-1}~ 
e^{-i\omega u} ~Y_{lm}(\theta,\phi)$ are purely \emph{outgoing} and 
($\hat{b}^{\dagger}_{\omega}$, $\hat{b}_{\omega}$), 
($\hat{c}^{\dagger}_{\omega}$, $\hat{c}_{\omega}$) are creation and annihilation 
operator pairs at \emph{future null infinity} $\scriplus$ and event 
horizon respectively. The solutions $\{{p}_{\omega}\}$ have zero Cauchy data on 
event horizon whereas the solutions $\{{q}_{\omega}\}$ have zero Cauchy data on 
future null infinity $\scriplus$.

\begin{figure}
\includegraphics[width=8.5cm]{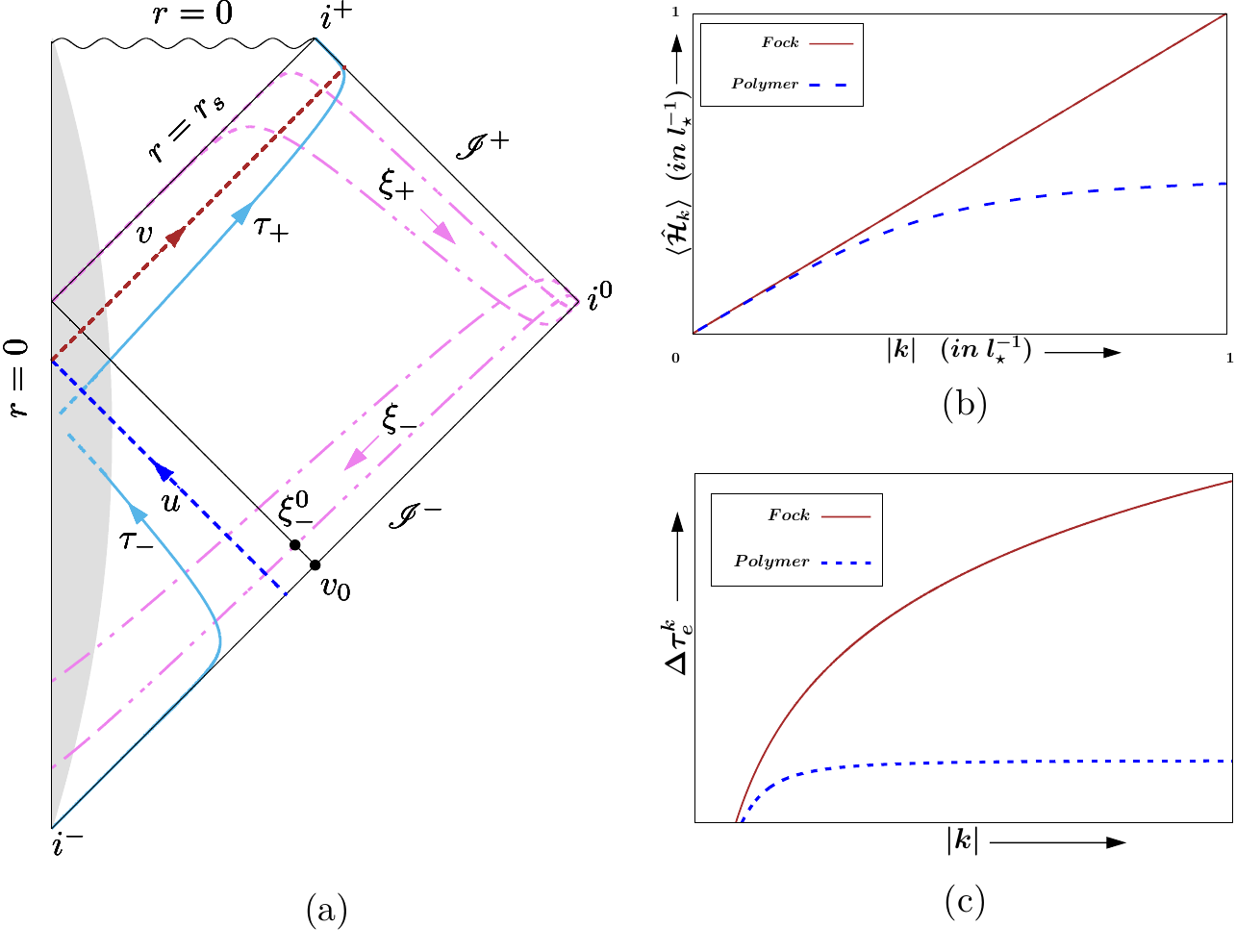}
\caption{(a) In Penrose diagram, shaded region represents the collapsing 
matter shell. Ingoing null rays  depart from past null infinity $\scriminus$ 
whereas outgoing null rays arrive at future null infinity $\scriplus$. 
Near-null coordinates are $\tau_{\pm}$ and $\xi_{\pm}$. (c) Arrival time 
$\Delta \tau_{e}^{k}$ for $k^{th}$ mode in arbitrary units.
}
\label{fig:NearNullPenrose} 
\end{figure}

\emph{Hawking radiation.}--
For derivation of Hawking effect, an essential relation between null coordinates 
on $\scriminus$ and $\scriplus$ with suitable choice of pivotal values (see 
FIG.\ref{fig:NearNullPenrose}) is given by
\begin{equation}\label{Relation:AdvancedRetardedNullCoordinates}
- u = - v  + 2\rs \ln \left(-v/2 \rs\right)  ~.
\end{equation}
The relation (\ref{Relation:AdvancedRetardedNullCoordinates}) crucially depends
on the fact that there was no black hole when relevant \emph{ingoing} modes 
departed from $\scriminus$. For Hawking radiation, relevant modes originate from 
the region $|v| \ll 2\rs$ on $\scriminus$ and for them the relation 
(\ref{Relation:AdvancedRetardedNullCoordinates}) can be approximated as
\begin{equation}\label{Relation:AdvancedRetardedNullCoordinatesApprox}
v \approx -2\rs~ e^{-u/2 \rs} ~~.
\end{equation}
The Hawking effect is realized from the expectation value of number operator 
corresponding to the observer at future null infinity $\scriplus$ in the vacuum 
state corresponding to the observer at past null infinity $\scriminus$, and is 
given by
\begin{equation}\label{NumberVEVHawking}
{N}_{\omega} \equiv 
\langle 0_{-}| \hat{b}^{\dagger}_{\omega} \hat{b}_{\omega} |0_{-}\rangle
= \frac{1}{e^{2\pi\omega/\ksg} - 1}  ~,
\end{equation}
where $\ksg=1/(2\rs)$ is the \emph{surface gravity} at the horizon. This 
perceived phenomena of blackbody radiation at $\scriplus$ is referred to
as the Hawking effect with Hawking temperature $T_H = \ksg/(2\pi k_B) = 
1/(8\pi G M k_B)$. We note that despite being mentioned frequently the 
derivation of the Hawking effect does not require any \emph{pair-production} of 
particles nor these particles would have any Cauchy data on $\scriminus$.

\emph{Canonical formulation.}--
In canonical formulation, field dynamics is viewed as `time evolution' of the 
modes on `spatial hypersurfaces'. So one needs to look beyond null coordinates 
as they do not lead to a true Hamiltonian that can describe evolution of the 
modes. We note that \emph{ingoing} field solutions (\ref{IngoingSolution}) have 
a phase factor $e^{-i\omega v}$. Along a given ingoing null trajectory advanced 
null coordinate $v$ is constant. However, one can use retarded null coordinate 
$u$ to parameterize its propagation. In other words, ingoing field solutions 
${f}_{\omega}(v)$, using the relation $v=u+2\rstar$, can be viewed as if 
${f}_{\omega}(u) = e^{-i\omega u } ~{f}_{\omega}(0)$ where $u$ varies along the 
trajectory. Remarkably, this form can be compared with time evolution of a 
Schrodinger wave function $\psi_{\omega}(\tau) = 
e^{-i\omega\tau}\psi_{\omega}(0)$ for a mechanical system with energy $\omega$ 
and time coordinate $\tau$. We also know that a massless, free scalar field can 
be mapped into a set of harmonic oscillators by using Fourier transformation. 
These insights then suggest that we may define a timelike coordinate by slightly 
deforming retarded null coordinate $u$ and define a  spacelike coordinate by 
deforming advanced null coordinate $v$ for an observer near past null infinity 
$\scriminus$, say $\observerminus$, as
\begin{equation}\label{NearNullCoordinatesMinus}
\tau_{-} = t - (1-\epsilon)\rstar ~~;~~ \xi_{-} = -t - (1+\epsilon)\rstar  ~,
\end{equation}
where $\epsilon$ is a real-valued parameter. In general, one can choose 
the parameter in the domain $0<\epsilon<2$ such that $\tau_{-}$ and $\xi_{-}$ 
are timelike and spacelike coordinates respectively. Here we choose the 
parameter $\epsilon$ to be a small and positive such that $\epsilon \gg 
\epsilon^2$. This choice of parameter allows us to mimic the basic tenets of 
the Hawking effect very closely. In any case, final result will be independent 
of the  explicit values of $\epsilon$.  Similarly, we define another set of 
timelike and spacelike coordinates $\tau_{+}$ and $\xi_{+}$ as
\begin{equation}\label{NearNullCoordinatesPlus}
\tau_{+} = t + (1-\epsilon)\rstar ~~;~~ \xi_{+} = -t + (1+\epsilon)\rstar  ~,
\end{equation}
for an observer near $\scriplus$, referred to as $\observerplus$. We note that 
one can algebraically transform the two sets of the coordinates 
(\ref{NearNullCoordinatesMinus},\ref{NearNullCoordinatesPlus}) to each other by 
simply substituting $\rstar \to -\rstar$. We have suitably chosen the 
directions of $\xi_{-}$ and $\xi_{+}$ for later convenience (See FIG. 
\ref{fig:NearNullPenrose}).

Similar to the relation (\ref{Relation:AdvancedRetardedNullCoordinates}), one 
can derive an analogous relation $\xi_{+}=\xi_{-} + 2 r_{s} 
\ln\left(\xi_{-}/2r_{s}\right)$ which can be approximated in the domain 
$|\xi_{-}| \ll 2r_{s}$, as
\begin{equation}\label{Relation:ximinusxiplus}
\xi_{-} \approx  2 \rs~e^{\xi_{+}/2 \rs} ~,
\end{equation}
where $\xi_{-}$ and $\xi_{+}$ refer to the spatial coordinates on a $\tau_{-} = 
constant$ and $\tau_{+} = constant$ surfaces for the observers $\observerminus$ 
and $\observerplus$ respectively  (details of the canonical derivation is 
provided in an accompanying paper \cite{Barman:2017a2}).


\emph{Scalar Field Hamiltonian.} --
The Hawking effect is crucially connected with the structure of the 
Schwarzschild metric in the $t-r$ plane. So for simplicity now onward we 
consider only the 1+1 dimensional system. For both the observers 
$\observerminus$ and $\observerplus$, the metrics are of the form
\begin{equation}\label{NearNullMetricMinus}
ds^2 = g^{\pm}_{\mu\nu}dx^{\mu}dx^{\nu} = \frac{\epsilon \,\Omega}{2}
\left[ - d\tau_{\pm}^2  + \frac{2}{\epsilon} d\tau_{\pm} d\xi_{\pm} 
+ d\xi_{\pm}^2 \right]  ~. 
\end{equation}
For large radial distances, the 4-dimensional scalar field action can be 
reduced to the form $S_{\varphi} =  \int d\tau_{\pm}  d\xi_{\pm} 
\left[-\tfrac{1}{2} \sqrt{-g^{0}} g^{0\mu\nu} \partial_{\mu}\varphi 
\partial_{\nu} \varphi \right]$, where $g^{\pm}_{\mu\nu} = (\epsilon \, 
\Omega/2) g^{0}_{\mu\nu}$. The metric $g^{0}_{\mu\nu}$ is flat but has 
off-diagonal terms. The corresponding scalar field Hamiltonians are
\begin{equation}\label{ScalarHamiltonianFullMinus}
H_{\varphi}^{\pm} = \int d\xi_{\pm}  \frac{1}{\epsilon}  \left[
\left\{ \frac{\Pi^2}{2}  + \frac{1}{2}  (\partial_{\xi_{\pm}}\varphi)^2 
\right\} + \Pi~ \partial_{\xi_{\pm}} \varphi \right] ~,
\end{equation}
where the \emph{lapse function} $N = 1/\epsilon$, the \emph{shift vector} $N^1 
= 
1/\epsilon$ and the determinant of the spatial metric $q=1$. The Poisson 
bracket 
between the field $\varphi$ and its conjugate momentum $\Pi$ can be written as
\begin{equation}\label{PoissonBracketMinus}
\{\varphi(\tau_{\pm},\xi_{\pm}), \Pi(\tau_{\pm},\xi_{\pm}')\} 
= \delta(\xi_{\pm} - \xi_{\pm}') ~.
\end{equation}
Using equations of motion, the field momentum can be expressed as 
$\Pi = \epsilon (\partial_{\tau_{\pm}}\varphi) - (\partial_{\xi_{\pm}}\varphi)$.

\emph{Fourier modes.}--
The Fourier modes of the scalar field can be defined for both observers
as
\begin{equation}\label{FourierModesDefinition}
\varphi =  \frac{1}{\sqrt{V_{\pm}}}\sum_{k} \tilde{\phi}_{k} e^{i k \xi_{\pm}} 
~;~ \Pi =  \frac{1}{\sqrt{V_{\pm}}} \sum_{k} \sqrt{q}~ \tilde{\pi}_{k} 
e^{i k \xi_{\pm}} ~,
\end{equation}
where $\tilde{\phi}_{k} = \tilde{\phi}_{k} (\tau_{\pm})$, 
$\tilde{\pi}_{k} = \tilde{\pi}_{k} (\tau_{\pm})$ are the complex-valued mode 
functions. The spatial volume $V_{\pm} = \int d\xi_{\pm}\sqrt{q}$ are formally 
divergent. To avoid dealing with explicitly divergent quantity, we choose a 
fiducial box with finite volume. Then the wave-vectors are $k \in \{k_r\}$ 
where 
$k_r = 2\pi r/L_{\pm}$ with $r$ being a non-zero integer and $L_{\pm}$ being 
the 
length of the box. The scalar field Hamiltonian 
(\ref{ScalarHamiltonianFullMinus}) can be expressed as $H_{\varphi}^{\pm} = 
\sum_k \tfrac{1}{\epsilon} (\mathcal{H}_k^{\pm} + \mathcal{D}_k^{\pm})$ where 
Hamiltonian density for $k^{th}$ mode $\mathcal{H}_k^{\pm} = \frac{1}{2} 
\tilde{\pi}_{k}  \tilde{\pi}_{-k} + \frac{1}{2} |k|^2 \tilde{\phi}_{k}  
\tilde{\phi}_{-k}$ and diffeomorphism generator $\mathcal{D}_k^{\pm} = - 
\frac{i k}{2} \left(\tilde{\pi}_{k} \tilde{\phi}_{-k} - \tilde{\pi}_{-k} 
\tilde{\phi}_{k} \right)$. 
The associated Poisson bracket is $\{\tilde{\phi}_{k}, \tilde{\pi}_{-k'}\} = 
\delta_{k,k'}$. We can relate the Fourier modes between the two different 
observers as
\begin{equation}\label{FieldModesRelation}
\tilde{\phi}_{\kr} = \sum_{k} \tilde{\phi}_{k} F_{0}(k,-\kr) ~,
\tilde{\pi}_{\kr} =  \sum_{k} \tilde{\pi}_{k} F_{1}(k,-\kr)
\end{equation}
where $\tilde{\phi}_{\kr} = \tilde{\phi}_{\kr}(\tau_{+}^0)$,  
$\tilde{\phi}_{k} = \tilde{\phi}_{k}(\tau_{-}^0)$, $\tilde{\pi}_{\kr} = 
\tilde{\pi}_{\kr}(\tau_{+}^0)$ and $\tilde{\pi}_{k} = 
\tilde{\pi}_{k}(\tau_{-}^0)$ \cite{Barman:2017a2}. 
The coefficient functions $F_{m}(k,\kr)$ are similar to the Bogoliubov 
coefficients in covariant formulation and are likewise formally divergent. It 
is 
shown \cite{Hossain:2014fma,Barman:2017a2} that these coefficients can be 
regularized to render them finite. The \emph{regulated} coefficients 
$F_{m}^{\delta}(\pm|k|,\kr)$ reduces to the exact expression when the regulator 
$\delta$ is removed \emph{i.e.} $\lim_{\delta\to0} F_{m}^{\delta}(k,\kr) = 
F_{m}(k,\kr)$ and satisfy following key relations \cite{Barman:2017a2}
\begin{eqnarray}\label{F0F0Relation}
F_{0}^{\delta}(-|k|,\kr) &=& e^{2\pi\rs\kr - i\delta\pi} 
~F_{0}^{\delta}(|k|,\kr)  ~, \nonumber \\
F_{1}^{\delta}(\pm|k|,\kr) &=& \mp \frac{\kr}{|k|}~F_{0}^{\delta}(\pm|k|,\kr) ~.
\end{eqnarray}

\emph{Number operator.}--
In order to quantize the scalar field we follow the method as used in 
\cite{Hossain:2010eb} where one canonically quantizes each Fourier mode. In 
particular, the expectation value of the Hamiltonian density operator  of a 
\emph{positive} frequency mode \emph{i.e.} $\kr>0$, for the observer 
$\observerplus$ in the vacuum state $|0_{-}\rangle$ of the observer 
$\observerminus$ \emph{i.e.}  $\langle \hat{\mathcal{H}}_{\kr}^{+} \rangle 
\equiv \langle 0_{-}| \hat{\mathcal{H}}_{\kr}^{+} |0_{-}\rangle$ can be 
expressed as \cite{Barman:2017a2}
\begin{equation}\label{HamiltonianPlusVEV}
\frac{\langle\hat{\mathcal{H}}_{\kr}^{+}\rangle}{\kr} =
\frac{e^{2\pi\kr/\ksg} + 1}{e^{2\pi\kr/\ksg} - 1} 
\left[ \frac{1}{\zeta(1+2\delta)}
\sum_{r=1}^{\infty} \frac{1}{r^{1+2\delta}} ~
\frac{\langle\hat{\mathcal{H}}_{k_r}^{-}\rangle}{k_r}
\right]
~,
\end{equation}
where \emph{Riemann zeta function} $\zeta(1+2\delta) = \sum_{r=1}^{\infty}
r^{-(1+2\delta)}$ and we have used the properties of the vacuum state such 
that $\langle 0_k|\hat{\phi}_{k} |0_k\rangle = 0$ and $\langle 
0_k|\hat{\pi}_{k} 
|0_k\rangle = 0$. 
We define the number density operator which represents the Hawking quanta as
\begin{equation}\label{NumberOperatorDefinition}
\hat{N}_{\kr} = \left[ 
\hat{\mathcal{H}}_{\kr}^{+} - \lim_{\ksg\to0} \hat{\mathcal{H}}_{\kr}^{+} 
\right] |\kr|^{-1}  ~,
\end{equation}
which makes it amply clear that existence of these quanta are tied to the 
non-zero values of the \emph{surface gravity} $\ksg$ at the horizon. Besides, 
this definition becomes crucial for the situation where the notion of creation 
and annihilation operators are not readily available like in polymer 
quantization.

The Fourier modes that we have considered so far, are in general complex valued 
functions. So in order to avoid double counting, as $\varphi$ is real-valued, 
here we make a \emph{choice} by setting imaginary components of 
the modes $\phi_{k}^i = 0$ and $\pi_{k}^i = 0$. This leads diffeomorphism 
generator $\mathcal{D}_k^{-}$  to vanish identically. Further, by 
redefining the real part of the modes as $\phi_{k} \equiv \phi_{k}^r$ and  
$\pi_{-k} \equiv \pi_{k}^r$, we can reduce the Hamiltonian density 
to its regular harmonic oscillator form $\mathcal{H}_k^{-} = \frac{1}{2} 
\pi_{k}^2 + \frac{1}{2} |k|^2\phi_{k}^2$ along with the Poisson bracket 
$\{\phi_{k}, \pi_{k'}\} = \delta_{k,k'}$ for the observer $\observerminus$.

\emph{Fock quantization.}--
The Fock vacuum state for the observer $\observerminus$ can be expressed as 
$|0_{-}\rangle = \prod_{k} \otimes |0_k\rangle$ where $|0_k\rangle$ is vacuum 
state of the $k^{th}$ oscillator. The corresponding energy spectrum can be 
written as $\hat{\mathcal{H}}_{k}^{-} |n_k\rangle = 
(n+\frac{1}{2})|k| |n_k\rangle$ where $n \ge 0$. Therefore, in Fock 
quantization $\langle\hat{\mathcal{H}}_{k}^{-}\rangle = \frac{1}{2}|k|$ for all 
modes. The Eqn. (\ref{HamiltonianPlusVEV}) and (\ref{NumberOperatorDefinition}) 
together then imply
\begin{equation}\label{NumberOperatorVEVFock}
N_{\omega} \equiv \langle \hat{N}_{\kr=\omega}\rangle = 
\frac{1}{e^{2\pi\omega/\ksg} 
- 1} = \frac{1}{e^{(4\pi\rs)\omega} - 1} ~,
\end{equation}
which corresponds to a thermal spectrum at Hawking temperature $T_H = 
\ksg/(2\pi k_B) = 1/(4\pi\rs k_B)$. It shows that we can derive the exact 
thermal spectrum of the Hawking effect also using Hamiltonian formulation.

\emph{Polymer quantization.}--
In polymer quantization, energy eigenvalues for the $k^{th}$ oscillator 
is given 
by \cite{Hossain:2010eb}
\begin{equation}\label{PolymerSHOSpectrumFull}
 \frac{E_{k}^{2n}}{|k|} = \frac{1}{4g} + \frac{g}{2} ~A_n(g)  ~,~
 \frac{E_{k}^{2n+1}}{|k|} = \frac{1}{4g} + \frac{g}{2} B_{n+1}(g) ~,
\end{equation}
where $n\ge0$, $A_n$ and $B_n$ are Mathieu characteristic value functions. The 
\emph{dimensionless} parameter $g \equiv |k|~l_{\star}$ where $\lstar$ is the 
\emph{polymer length scale}. For small $g$, the energy spectrum 
(\ref{PolymerSHOSpectrumFull}) reduces to $ E_{k}^{2n}/|k| \approx 
E_{k}^{2n+1}/|k| = \left( n +\tfrac{1}{2}\right) + \mathcal{O}(g)$. It implies 
that polymer quantization correctly reproduces the spectrum for sub-Planckian 
modes. However, significant non-perturbative modifications in the spectrum are 
seen for super-Planckian modes. In particular, for large $g$, ground state 
energy can be approximated as $E_{k}^0/|k| = 1/4g + \mathcal{O}(g^{-3})$. So 
unlike in Fock quantization where $\langle\hat{\mathcal{H}}_{k}^{-}\rangle/k = 
\tfrac{1}{2}$ for all $k$, in polymer quantization 
$\langle\hat{\mathcal{H}}_{k}^{-}\rangle/k \to 0$ for the trans-Planckian modes 
as $k\to\infty$. Therefore, when one removes the regulator $\delta$ in polymer 
quantization, the expectation value of the number operator 
(\ref{NumberOperatorDefinition}), due to the form of Eqn. 
(\ref{HamiltonianPlusVEV}) and the \emph{zeta function identity} 
$\lim_{s\to0}[s~\zeta(1+s)]=1$, becomes 
\begin{equation}\label{NumberOperatorVEV}
N_{\omega}^{poly} = \langle \hat{N}_{\omega}^{poly}\rangle = 0 ~.
\end{equation}
This property of the number operator, having same mathematical expression, is 
identical to the case of Unruh effect \cite{Hossain:2014fma}. Therefore, an 
asymptotic future observer $\observerplus$ would not perceive any Hawking 
quanta in polymer quantization, in contrary to the Fock quantization.

\emph{Duration of Hawking effect.}--
We note that expectation value of number operator is \emph{discontinuous} in 
the limit $\lstar \to 0$. In order to physically understand this behavior, 
let us consider a future observer located at a fixed $r \gg \rs$. The proper 
time interval $\Delta \tau$ for this observer then follows the relation $\Delta 
\tau = \Delta t =  \Delta \tau_{+}$, as fixed $r$ implies $\Delta \tau_{+} = 
-\Delta \xi_{+}$. For this observer, the difference in arrival time for two 
modes which were emitted from the coordinate points $\xi_{-}^{1}$ and 
$\xi_{-}^{2}$ on a fixed $\tau_{-}$ surface near $\scriminus$ can be written 
using the relation (\ref{Relation:ximinusxiplus}) as $\Delta \tau \approx 
2r_{s}\ln\left(\xi_{-}^{1}/\xi_{-}^{2}\right)$. The relevant modes for 
Hawking radiation are emitted from the region $|\xi_{-}| \ll 2r_{s}$. So for
simplicity we choose the arrival time of the mode emitted from 
$\xi_{-}^{1}=2\rs$ as the `beginning' of Hawking radiation. Then the difference 
in arrival time for a mode emitted from a general coordinate point 
$\xi_{-}^{e}$ ($\le 2r_{s}$) would be
\begin{equation}\label{GeneralArrivalTime}
\Delta \tau_{e} = 2r_{s}\ln\left(2\rs/\xi_{-}^{e}\right)  ~.
\end{equation}
Clearly, the modes whose point  of emission $\xi_{-}^{e}$ are closer to the 
origin, arrive later near $\scriplus$. For a relativistic mode in ground state, 
one can always associate a de-Broglie wavelength (like `width' of the quanta) 
as $\lambda_k^{de} = h c/E^0_k = h c/\langle\hat{\mathcal{H}}_{k}^{-}\rangle$. 
Naturally, point of emission cannot be made more accurate than its de-Broglie 
wavelength. Hence we choose closest possible point of emission for the $k^{th}$ 
mode as $\xi_{-}^{e} = \lambda_k^{de}$. 
It may be checked that the proper wavelength of these modes at the time of 
arrival near $\scriplus$ would be $\lambda_k^{out} \sim 2\rs$ which 
can be viewed as the \emph{Wien's displacement law} for Hawking radiation.
We now define the `duration' of the Hawking effect to be the arrival time of 
the mode with least possible de-Broglie wavelength, given by
\begin{equation}\label{HawkingDuration}
\tau^{H} \equiv  max( \{\Delta \tau_{e}\}) = \lim_{k\to\infty} 2r_{s} \ln 
\left[ \frac{\rs}{\pi} ~\langle\hat{\mathcal{H}}_{k}^{-}\rangle \right]  ~.
\end{equation}
In Fock quantization $\langle\hat{\mathcal{H}}_{k}^{-}\rangle = \frac{1}{2}|k|$ 
for all modes, including trans-Planckian modes, which implies $\tau^{H} \to 
\infty$. On the contrary, for trans-Planckian modes in polymer quantization  
$\langle\hat{\mathcal{H}}_{k}^{-}\rangle \approx \tfrac{1}{4\lstar}$ which 
implies $\tau^{H} \approx 2\rs \ln \left(\rs/4\pi\lstar \right)$
(see FIG.\ref{fig:NearNullPenrose}). 
If we take $\lstar$ to be Planck length then for a solar mass black hole this 
duration $\tau^{H} \approx 1.7$ \emph{milliseconds} whereas for an 
ultra-massive black hole with mass $M = 4\times10^{10} M_{\odot}$ (like one at 
the center of the galaxy \emph{S5 0014+81}) the duration $\tau^{H} \approx 2.8$ 
\emph{years}. This short duration of the Hawking effect explains why an 
observer in asymptotic future would not perceive any Hawking quanta in polymer 
quantization.

\emph{Information loss paradox.}--
In Fock quantization, the Hawking effect persists \emph{ad infinitum}. 
Therefore, one argues that it would eventually lead to a complete evaporation of 
the black hole. This in turn leads to the so-called information loss paradox 
\cite{Information:3article,Unruh:2017uaw}, as from thermal radiation alone one 
cannot recover information about the collapsing matter shell which led to the 
formation of the black hole. Recently Unruh and Wald have classified the 
proposals to resolve information loss paradox in four categories: (I) 
\emph{fuzzball} formation (II) \emph{firewall}  scenario (III) Planckian 
\emph{remnant} and (IV) Planckian \emph{final burst} \cite{Unruh:2017uaw}. 
However, in the scenario as implied by the polymer quantization, the Hawking 
radiation stops after a short duration leaving the classical black hole 
unchanged. Consequently, there is no loss of information. Therefore, it provides 
a new way to resolve the information loss paradox and it requires modification 
only in the trans-Planckian physics.

\emph{Discussions.}--
In summary, we have shown that the  Hawking effect is short-lived in polymer 
quantization of matter field and it eventually disappears to an asymptotic 
future observer. In order to arrive at these results we have introduced a set of 
near-null coordinates which allowed us to have an exact derivation of the 
Hawking effect using Hamiltonian formulation. In polymer quantization, the 
duration of the Hawking effect would appear to be few milliseconds for a solar 
mass black hole whereas it would be few years for an ultra-massive black hole. 
These predictions are testable in principle and may allow one to verify or rule 
out the given hypothesis. Furthermore, this short-lived Hawking effect scenario 
provides a new way to resolve the so-called information loss paradox.

\emph{Acknowledgments.}--
We thank Gopal Sardar for discussions. SB and CS thank IISER 
Kolkata for supporting this work through doctoral fellowships.

\end{document}